\documentclass[sigconf,screen]{acmart}

\usepackage{blindtext}

\usepackage{booktabs} 
\usepackage{balance}
\graphicspath{ {Figures/} }
\usepackage{tabularx}
\usepackage{float}
\usepackage{amsmath}
\usepackage{enumitem}

\usepackage{multirow}
\usepackage{color}

\usepackage{pgfplots}
\pgfplotsset{compat=1.8}
\usepackage{filecontents} 

\usepackage{listings}

\usepackage{tikz}
\usetikzlibrary{calc}
\usetikzlibrary{trees}
\usetikzlibrary{arrows}
\usetikzlibrary{graphs}
\usetikzlibrary{positioning}
\usetikzlibrary{shapes.geometric}
\usetikzlibrary{shapes.misc}
\tikzstyle{block}=[draw opacity=0.7,line width=1.0cm]

\tikzset{
modal/.style={>=stealth',shorten >=1pt,shorten <=1pt,auto,node distance=1.5cm, semithick},
world/.style={circle,draw,minimum size=0.5cm,fill=gray!15},
ghost/.style={circle,draw,minimum size=0.5cm,fill=white,draw=white, fill opacity=0},
atom/.style={rectangle,minimum size=0.5cm,fill=green!50!black!20,draw=green!50!black},
structure/.style={=>stealth’,shorten >=1pt,shorten <=1pt,auto,node distance=25mm, thick, minimum size=5mm},
set/.style={rectangle, minimum size=0.5cm,fill=green!50!black!20,draw=green!50!black, font=\sffamily\scriptsize},
lattice/.style={=>stealth’,shorten >=1pt,shorten <=1pt, auto, node distance=1cm, semithick, minimum size=3mm},
every node/.style={font=\sffamily\footnotesize},
point/.style={circle,draw,inner sep=0.5mm,fill=black},
scriptsize/.style={font=\sffamily\scriptsize},
tiny/.style={font=\sffamily\tiny},
highlight/.style={fill=red!50!black!20,draw=red!50!black},
comment/.style={rectangle, fill=white,text opacity=1,fill opacity=0, draw opacity=0, minimum size=0.5cm,font=\itshape\sffamily\tiny},
reflexive above/.style={->,loop,looseness=7,in=120,out=60},
reflexive below/.style={->,loop,looseness=7,in=240,out=300},
reflexive left/.style={->,loop,looseness=7,in=150,out=210},
reflexive right/.style={->,loop,looseness=7,in=30,out=330}
}

\usepackage{color}

\newcommand{\AIE}{AlloyInEcore}
\newcommand{\E}{ECAS}

\usepackage{chngcntr}
\counterwithin*{equation}{subsection}

\setcopyright{acmcopyright}
\acmPrice{15.00}
\acmDOI{10.1145/3236024.3264588}
\acmYear{2018}
\copyrightyear{2018}
\acmISBN{978-1-4503-5573-5/18/11}
\acmConference[ESEC/FSE '18]{Proceedings of the 26th ACM Joint European Software Engineering Conference and Symposium on the Foundations of Software Engineering}{November 4--9, 2018}{Lake Buena Vista, FL, USA}
\acmBooktitle{Proceedings of the 26th ACM Joint European Software Engineering Conference and Symposium on the Foundations of Software Engineering (ESEC/FSE '18), November 4--9, 2018, Lake Buena Vista, FL, USA}

\begin{document}
\title[AlloyInEcore: Embedding of First-Order Relational Logic into ...]{AlloyInEcore: Embedding of First-Order Relational Logic into Meta-Object Facility for Automated Model Reasoning}

\author{Ferhat Erata}
\affiliation{%
  \institution{UNIT Information Technologies R\&D \country{Turkey}}
}


\author{Arda Goknil}
\affiliation{%
  \institution{University of Luxembourg \country{Luxembourg}}
}

\author{Ivan Kurtev}
\affiliation{%
  \institution{Altran Netherlands \country{the Netherlands}} 
}

\author{Bedir Tekinerdogan}
\affiliation{%
  \institution{Wageningen University \country{the Netherlands}} 
}


\begin{abstract}

We present \AIE, a tool for specifying metamodels with their static semantics to facilitate automated, formal reasoning on models. Software development projects require that software systems be specified in various models (e.g., requirements models, architecture models, test models, and source code). It is crucial to reason about those models to ensure the correct and complete system specifications. \AIE~allows the user to specify metamodels with their static semantics, while, using the semantics, it automatically detects inconsistent models, and completes partial models. It has been evaluated on three industrial case studies in the automotive domain (\url{https://modelwriter.github.io/AlloyInEcore/}).

\end{abstract}

%
%



\maketitle

\section{Introduction} 
\label{sec:intro}

Model Driven Engineering (MDE) is becoming a crucial practice in industry due to the increasing complexity of software systems that warrant better support for managing development artifacts~\cite{hutchinson2011empirical}. 
In MDE, software is developed by successively transforming abstract models to more concrete ones. Each model conforms to its metamodel, an artefact usually created using Ecore~\cite{Ecore}, a de facto industry standard for metamodeling and an example of implementation of the Meta-Object Facility (MOF)~\cite{MOF} which describes the means to create and manipulate models and metamodels. An important challenge in MDE is providing ability of automated, formal reasoning on models, e.g., checking model consistency and completing partial models~\cite{france2007model,SemanticTrace2011}. 

We present a tool, \AIE, which allows specification of metamodels with their static semantics and facilitates multiple forms of automated, formal reasoning on models. \AIE~is targeted at environments that require integration and reasoning on heterogeneous models. Such environments are often encountered within the context of our research~\cite{ModelWriter, Assume} in collaboration with Ford-Otosan~\cite{FordOtosan}. 
The key idea behind \AIE~is that the static semantics of an Ecore metamodel can be specified within a simple first-order logic of sets and relations to support reasoning on models conforming to the metamodel. 

Alloy~\cite{AlloyBook} is a declarative modeling language based on first-order relational logic. It has been explored by the MDE community for the purpose of analyzing UML/OCL models~\cite{anastasakis2007uml2alloy, shah2009uml2Alloy, cunha2015UML2Alloy}. 
Most of the existing tools and approaches use a transformation of UML/OCL models to Alloy, which, however, does not support directly some important concepts like multiple inheritance, generic types, and type parameters due to the fundamental differences between UML/OCL and Alloy notations~\cite{cunha2015UML2Alloy, vaziri2000some}. In the case of dealing with various models in different abstraction levels, it is required to enable the specification of such concepts, and herewith multiple forms of model reasoning. To do so, \AIE~provides the following major features: (i) Alloy-like notation embedded into Ecore to specify the static semantics of metamodels based on First-order Logic (FOL), relational operators, and transitive closure; (ii) direct translation of Ecore metamodels into the language of Kodkod~\cite{torlak2007KodKodModelFinder}, 
an efficient SAT-based constraint solver for FOL with relational algebra. In this way we avoid the problems related to the differences between Alloy and Ecore. Our tool performs two major model reasoning tasks: completing partial models and detecting inconsistent model parts.

\section{Related Work}
\label{sec:related}

Several formal analysis methods and 
specification languages have been proposed relying on modern SAT-solvers, SMT solvers and theorem provers (e.g., Formula \cite{jackson2011Formula} using Z3 SMT-solver \cite{z3}, Clafer~\cite{kacper2016clafer} using Alloy along with Choco CSP solver~\cite{choco}, and Alloy using KodKod \cite{torlak2007KodKodModelFinder} that relies on SAT solvers like Minisat \cite{Minisat}). 



A number of MDE solutions and tools provide automated model reasoning using existing formal analysis methods and specification languages based on constraint logic programming (e.g.,~\cite{cabot2008verificationUMLusingCSP, cabot2007umltocsp}), SAT-based model finders (e.g.,~\cite{anastasakis2007uml2alloy, shah2009uml2Alloy, soeken2010satUSE, 
cunha2015UML2Alloy}), and SMT solvers (e.g.,~\cite{ElGhazi2011Alloy2SMT, use2016smt}). However, to the best of our knowledge, none of them provides a method that embeds FOL augmented with relational calculus into MOF/Ecore to specify the static metamodel semantics with the support for partial models, composition, cardinality constraints, multiple inheritance, generic types, and type parameters. For instance, Anastakis et al.~\cite{anastasakis2007uml2alloy} advocate to transform UML/OCL specifications to Alloy for automated model reasoning. The proposed transformation does not support multiple inheritance, generic types, and type parameters because of the differences between UML/OCL and Alloy notations. Lightning~\cite{Lightning} is a tool-supported approach for defining some aspects of Domain-Specific Languages (e.g. abstract syntax and semantics) entirely in Alloy. In our approach, language designers can use \AIE~to specify the abstract syntax of a language as an Ecore metamodel enriched with embedded Alloy-like  statements.  USE~\cite{USE2KodKod2012} is a tool for analyzing models expressed in UML and OCL. Similarly to our approach, USE translates models into relational logic and relies on the Kodkod library.

\section{Tool Overview} \label{sec:overview}

\begin{figure}[ht]
\centering
\vspace*{-1.2em}
\includegraphics[width=0.90\linewidth]{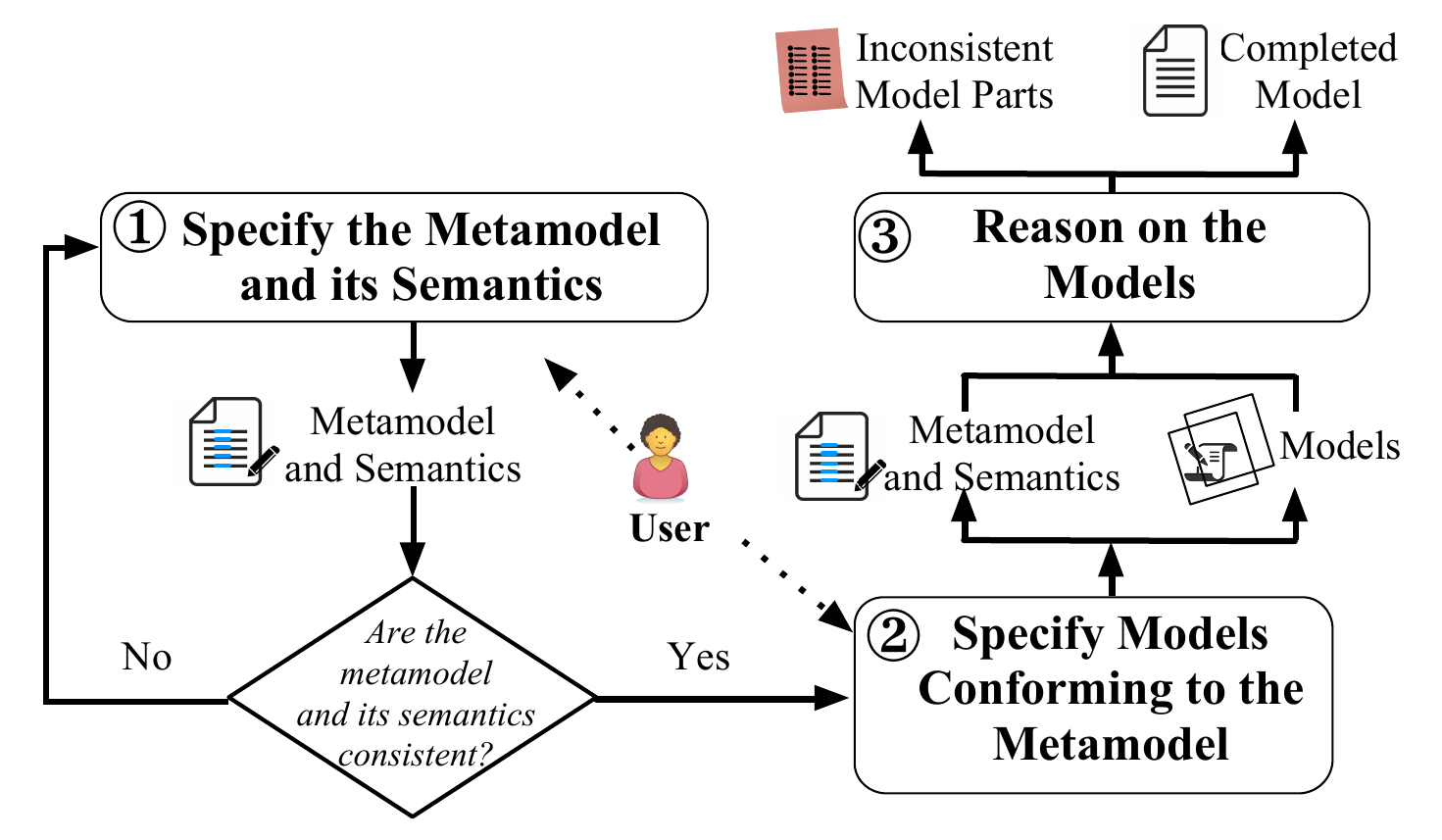}
\vspace*{-1.2em}
\caption{Tool Overview}\label{fig:tool_overview}
\vspace*{-1.0em}
\end{figure}

\AIE~is a generalization of our previous tool Tarski \cite{erata2017tool, ferhat2017SAC:PL}, which reasons on trace links in traceability models using configurable trace link semantics. Fig.~\ref{fig:tool_overview} presents an overview of our tool. 
In Step 1, the user specifies an Ecore metamodel and its static semantics expressed in FOL augmented with 
the relational calculus~\cite{tarski1941calculus} embedded in Ecore. To do so, \AIE~ natively supports \textit{Alloy} \textit{in} \textit{Ecore} with a custom Eclipse editor.  

Once the user specifies the metamodel and its semantics, \AIE~ allows creation of instance model(s) 
conforming to the metamodel (Step 2).
After the model is created, the tool proceeds to Step 3 using automated model reasoning. In the following, we elaborate each step 
using the \textit{theory of lists} as a running example. 

\subsection{Specification of Metamodels and Semantics} 
\label{subsec:specification}
\begin{figure}[htbp]
\centering
\includegraphics[width=0.87\linewidth]{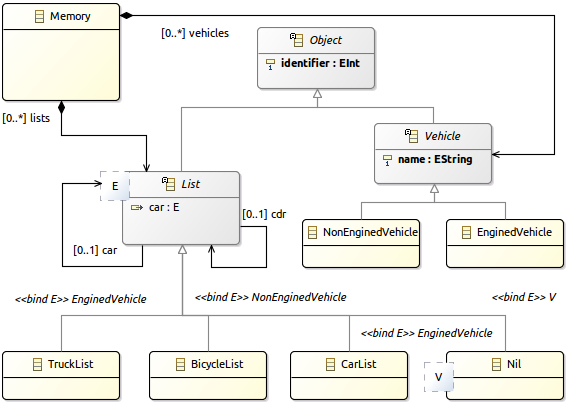}
\vspace*{-1.2em}
\caption{The Metamodel for the Theory of Lists}\label{fig:class_diagram}
\vspace*{-1.7em}
\end{figure}

As the first step, the user specifies a metamodel and its semantics in our Alloy-like notation embedded into Ecore. 
The user can create the metamodel using any graphical, textual, or tree-based Ecore model editor including our \AIE~editor. Fig.~\ref{fig:class_diagram} gives the \textit{theory of lists} metamodel in the Ecore graphical editor.

\begin{figure}[ht]
\centering
\vspace*{-1.0em}
\includegraphics[width=0.98\linewidth]{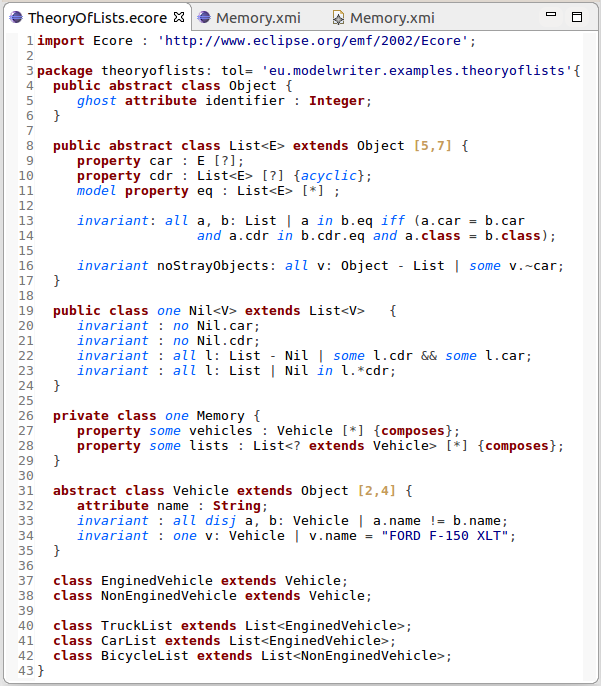}
\vspace*{-1.2em}
\caption{Example Metamodel Semantics in \AIE}\label{fig:spec_tol}
\vspace*{-1.0em}
\end{figure}

The user uses our editor to specify the static semantics of the metamodel (see Fig.~\ref{fig:spec_tol}). The keywords in blue and brown are introduced by \AIE~for specifying the metamodel semantics, while the ones in red are the Ecore keywords for defining the metamodel itself. Some of the \AIE~keywords (e.g., \textit{ghost}, \textit{model}, and \textit{nullable}) were borrowed from JML~\cite{JML2005}.

There are three abstract classes in Fig.~\ref{fig:spec_tol}: \textit{Object}, \textit{List}, and \textit{Vehicle}. The \textit{Object} is an abstract class at the root of the class hierarchy. The \textit{ghost} keyword indicates that the \textit{identifier} attribute will not be considered in the model reasoning (Line 5). The cardinality constraint in Line 8 specifies the lower and upper bounds on the \textit{List} instances. The abstract class \textit{List} is composed of two properties: the \textit{car} mapping each \textit{List} instance to an instance of another class (e.g., a \textit{Vehicle} instance) and the \textit{cdr} pointing to another \textit{List} instance. 
The \textit{?} keyword constrains these properties to be partial functions (Lines 9-10). To rule out cyclic lists, the \textit{acyclic} keyword is used in the \textit{cdr} property (Line 10). 
The \textit{model} keyword defines the \textit{eq} property as a relation to be inferred in the reasoning (Line 11). \textit{Model} properties are not mapped to Ecore features. 

Metamodel semantics is mainly given as \textit{invariants}. For instance, the invariant in Lines 13-14 ensures 
that two \textit{List} instances are equal (\textit{'a in b.eq'} in Line 13) if and only if the head instances in the two lists are the same (\textit{'a.car = b.car'} in Line 13), the subsequent \textit{List} instances are equal (\textit{'a.cdr in b.cdr.eq'} in Line 14), and the \textit{List} instances are of the same type (\textit{'a.class = b.class'} in Line 14). 
The invariant in Line 16 guarantees that each \textit{Vehicle} instance is in at least one list (see the \textit{some} keyword).
The \textit{Nil} class represents the empty list (Line 19). The \textit{one} keyword makes the \textit{Nil} class a singleton set, which means there can be only one \textit{Nil} instance in a model. A \textit{Nil} instance has neither the \textit{car} nor the \textit{cdr} property (Lines 20-21), while a Non-nil \textit{List} instance has both \textit{car} and \textit{cdr} (Line 22). A \textit{Nil} instance is always a subsequent list of any \textit{List} instance (Line 23). 
The singleton class \textit{Memory} holds the \textit{Vehicle} and \textit{List} classes (see the \texttt{composes} keyword in Lines 27-28). 
It is important to note that `\textit{{List<? extends Vehicle>}}' represents the \textit{List} instances of any subclass of \textit{Vehicle} (Line 28). Each \textit{Vehicle} instance has a unique name (Line 33) and there is always exactly one \textit{Vehicle} instance with the name "Ford F-150 XLT" (Line 34). There are two types of vehicles: \textit{NonEnginedVehicle} and \textit{EnginedVehicle} (Line 37-38). \textit{TruckList} and \textit{CarList} are lists of \textit{EnginedVehicles} (Lines 40-41), while \textit{BicycleList} is a list of \textit{NonEnginedVehicle} (Line 42).

\subsection{Specification of Models} \label{subsec:models}

The user can use any graphical, textual, or tree-based Ecore model editor to specify models conforming to the metamodel (Step 2 in Fig.~\ref{fig:tool_overview}). Before creating any model, \AIE~automatically checks if the user can specify at least one valid model that conforms to the metamodel and its static semantics. The user may have specified some contradicting invariants where it is not possible to create a valid model. \AIE~automatically identifies the contradictions in the metamodel specification and notifies the user.


\subsection{Automated Reasoning on Models} \label{subsec:reasoning}
Model completion and consistency checking aim at deriving new instances and relations in the given model, and determining model parts violating the metamodel semantics, respectively. These two activities are processed as a single reasoning activity because they use
the same reasoning machinery. The consistency checking can be considered as part of model completion because a partial model is completed only if it is consistent.

\vspace*{-0.5em}

\subsubsection{Checking Model Consistency} 
\label{subsubsec:consistency}
\begin{figure}[tb]
\centering
\includegraphics[width=0.91\linewidth]{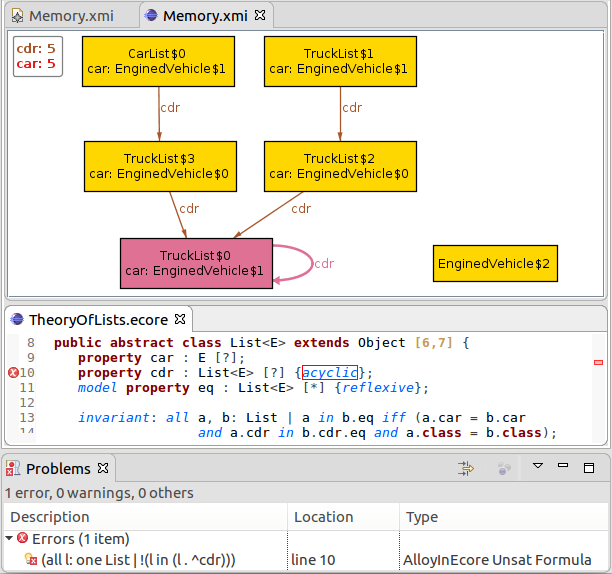}
\vspace*{-1.1em}
\caption{Example Inconsistent Model of the Theory of Lists}\label{fig:inconsistency_detected}
\vspace*{-1.75em}
\end{figure}

\AIE~takes a model and its metamodel as input, and automatically identifies, using the static metamodel semantics, inconsistent model parts as output. \AIE~provides an explanation of the inconsistency by giving all the instances and relations causing the inconsistency. Fig.~\ref{fig:inconsistency_detected} gives three \AIE~panes for an example inconsistent model of the theory of lists. The first pane in Fig.~\ref{fig:inconsistency_detected} gives the inconsistent model, while, in the second pane, \AIE~highlights part of the metamodel semantics causing the inconsistency. Although the \textit{cdr} property of the \textit{List} class is given \textit{acyclic} (see the highlighted \textit{acyclic} keyword in Line 10), the red colored \textit{cdr} in \textit{TruckList\$0} is referring to the instance itself. The third pane in Fig.~\ref{fig:inconsistency_detected} gives the first order relational logic formula that corresponds to the \textit{acyclic} keyword for further explanation of the inconsistency.

\vspace*{-0.5em}
\subsubsection{Completing Partial Models} 
\label{subsubsec:completing}
\begin{figure}[htbp]
\centering
\vspace*{-0.5em}
\includegraphics[width=0.91\linewidth]{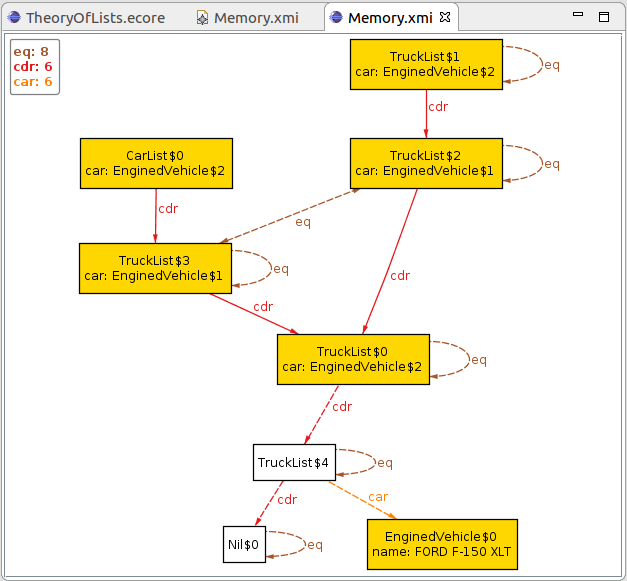}
\vspace*{-1.1em}
\caption{Example Completed Model of the Theory of Lists}\label{fig:reasoning_result}
\vspace*{-1.5em}
\end{figure}

If the given model is consistent, \AIE~automatically deduces new instances and relations in the input model using the static metamodel semantics. The model is completed only if it is consistent and not an exact model (i.e., a model where it is not possible to infer anything more). For instance, the user removes the cyclic \textit{cdr} relation in Fig.~\ref{fig:inconsistency_detected} to make the model consistent, and then \AIE~completes the model. Our tool infers 14 different complete models for the partial model in Fig.~\ref{fig:inconsistency_detected}. Fig.~\ref{fig:reasoning_result} shows one of these inferred models.

The dashed arrows and the white boxes represent the automatically inferred relations and instances, respectively. \AIE~automatically infers one \textit{Nil} instance (see the \textit{one} keyword in Line 19 in Fig~\ref{fig:spec_tol}) and one \textit{TruckList} instance (see the invariant in Line 16 and the cardinality constraint in Line 8). It also infers the name for \textit{EnginedVehicle\$0} (see the invariant in Line 34) and the \textit{car} relation between \textit{TruckList\$4} and \textit{EnginedVehicle\$0} (see the invariant in Line 16). The inferred \textit{Nil} instance is a subsequent list of the inferred \textit{TruckList} instance (see the invariant in Line 23). Each \textit{List} instance is equal to itself, while \textit{TruckList\$3} is equal to \textit{TruckList\$2} (see the invariant in Lines 13-14).

\section{Evaluation} 
\label{sec:evaluation}

Our goal was to assess, in an industrial context, the benefits of using \AIE~to facilitate automated model reasoning using user-defined metamodel semantics. 
We selected three case studies from the Electronically Controlled Air Suspension (\E) system developed by Ford-Otosan~\cite{FordOtosan}. Each case study is with an \E~artifact conforming to a different metamodel (i.e., a requirements specification, a data flow diagram and a SysML model).

Before conducting the case studies, the Ford-Otosan engineers were given presentations illustrating the \AIE~steps and a tool demo. The engineers held various roles (e.g., senior software and system engineers) with substantial development experience.
For each case study, we assisted the engineers in specifying metamodel semantics in \AIE~(the 1st, 2nd and 3rd columns in Table~\ref{tab:usecases}).

\begin{table}[ht] 
 \vspace*{-1.2em}
\caption{Number of Classes, Properties, Invariants, Models and Completed \& Inconsistent Parts in the Case Studies} \label{tab:usecases}
 \vspace*{-1.0em}
   \begin{tabular}{@{}c c c c  c c c@{}}
    \toprule
       \multicolumn{1}{p{0.03\columnwidth}}{\centering } 
     & \multicolumn{1}{p{0.10\columnwidth}}{\centering Meta-\\classes} 
     & \multicolumn{1}{p{0.10\columnwidth}}{\centering Pro-\\perties}  
     & \multicolumn{1}{p{0.10\columnwidth}}{\centering Invar-\\iants}   
     & \multicolumn{1}{p{0.13\columnwidth}}{\centering Model\\Elements}  
     & \multicolumn{1}{p{0.13\columnwidth}}{\centering Comple.\\Elements}
     & \multicolumn{1}{p{0.135\columnwidth}}{\centering Inconsis.\\Elements}\\
    \midrule
    $\#1$ &3 &7 &42 &116 &480 &5\\ 
    $\#2$ &5 &8 &6 &51 &114 &1\\ 
    $\#3$ &15 &12 &10 &135 &432 &2\\ 
    \bottomrule
 \end{tabular}
 \vspace*{-1.0em}
\end{table}

To evaluate the output, we had semi-structured interviews with the engineers. All the completed model parts and the identified inconsistencies were confirmed by the engineers to be correct (the 4th and 5th columns). The engineers considered the automated reasoning on models to be highly valuable. The Alloy-like notation embedded into Ecore was sufficient and easy for the engineers to specify the metamodel semantics in the case studies. For the largest model (the 3rd row), it took 126 secs to perform the reasoning.

\section{Implementation \& Availability} 
\label{sec:impl}

\AIE~has been implemented as an Eclipse plug-in. 
We use Kodkod~\cite{torlak2007KodKodModelFinder} 
to perform automated reasoning on models based on the metamodel semantics. 
\AIE~translates the input metamodel and semantic specification into a first order relational formula. It also translates the input model into a \textit{Universe} and \textit{Bounds} in KodKod. 
Kodkod translates the formula and the bounds into a Boolean satisfiability (SAT) problem to invoke an off-the-shelf SAT solver. If the SAT solver finds a SAT solution to the problem, Kodkod translates that SAT solution into a solution to the formula from which \AIE~derives the completed model.

\AIE~is approximately 31K lines of Java code, excluding comments and third-party libraries.
Additional details about \AIE, including executable files and a screencast covering motivations, are available on the tool's website at: 
\begin{center}
\fbox{\bf
{\url{https://modelwriter.github.io/AlloyInEcore/}}}
\end{center}

\section{Conclusion} 
\label{sec:conclusion}

We presented a tool that enables the specification of metamodel semantics for automated model reasoning. 
The key characteristics of our tool are (1) enabling the user to specify metamodels and their semantics in a single environment, (2) identifying inconsistent model parts, and (3) completing partial models. The tool has been evaluated over three industrial case studies. 

\begin{acks}
This work is conducted within ASSUME~\cite{Assume} project and partially supported by the Scientific and Technological Research Council of Turkey under project \#9150181, and by the European Research Council under the European Union's Horizon 2020 research and innovation programme (grant no 694277). 
\end{acks}


\bibliographystyle{ACM-Reference-Format}
\balance
\bibliography{sigproc,literature}

\section*{Appendices}

We provide three appendices for the paper. In Appendix \ref{apx:availability}, we provide source code, tool and data availability details. In Appendix \ref{apx:grammar}, we provide the simplified and readable version of the grammar of the \AIE's recognizer. Finally in Appendix \ref{apx:translation}, we present the log file representing the full translation of the use case, Theory of List, running on the \AIE~. 

\appendix
\nobalance

\section{Availability \& Open Source License} \label{apx:availability} 

\noindent\textbf{Source Codes, Screencast and Datasets}. The source codes files and datasets of \AIE~are publicly available for download and use at the project website. A screencast and the installation steps for \AIE~are also available at the same website and can be found at: 

\begin{center}
\fbox{\bf{\url{https://modelwriter.github.io/AlloyInEcore/}}}
\end{center}

\noindent\AIE~ is being developed under Work Package 3 and 5 within ASSUME~\cite{Assume} project, labeled by the European Union's EUREKA Cluster programme ITEA (Information Technology for European Advancement). Further details about the project can be found at: 

\begin{center}
\fbox{\bf{\url{https://itea3.org/project/assume.html}}}
\end{center}

\section{The Grammar of the Front-end}
\label{apx:grammar}

The simplified version of the grammar that parses AlloyInEcore models is presented in the following.

    \begin{flalign*}
    package &\Coloneqq \mathbf{package} \ identifier \ \{ \ package^{\ast} \ classifier^{\ast} \ invariant^{\ast} \ \} &\\
    classifier &\Coloneqq class \mid dataType \mid enum &\\
    class    &\Coloneqq \mathbf{abstract}? \ cardinality? \ \mathbf{class} \ identifier \ template &\\
             &\quad \; \ \ \mathbf{extends} \ type^{\ast}  \ bound? \ \{ \ feature^{\ast}  \ invariant^{\ast} \ \} &\\
    feature  &\Coloneqq  attribute \mid reference &\\
    attribute &\Coloneqq qualifier^{\ast} \ \mathbf{attribute} \ cardinality? \ identifier &\\
              &\quad \; : \ type \ multiplicity \ \{ \ \mathbf{id}? \ \mathbf{derived}? \ \} &\\
    reference &\Coloneqq qualifier^{\ast} \ \mathbf{property} \ cardinality? \ identifier &\\
              &\quad \; : \ type \ multiplicity \ \{ \ \mathbf{derived}? \ \mathbf{composes}? \ props^{\ast} \ \} &\\
    mult &\Coloneqq  [ \ constant \ (.. \ constant)? \mid ({}*{} | {}+{} | {}?{}) \ ] &\\
    type &\Coloneqq identifier \ (< \ argument \ (, \ argument)^{\ast} \ >)? &\\
    argument &\Coloneqq  type \mid  wildcard &\\
    wildcard &\Coloneqq \ ? \ ((\mathbf{extends} \mid \mathbf{super}) \ type))? &\\
    template &\Coloneqq {}<{} parameter \ (, \ parameter)^{\ast} > &\\
    parameter &\Coloneqq identifier \ (\mathbf{extends} \ type \ (\& \ type)^{\ast} )? &\\
    bound &\Coloneqq [ \ constant, \ constant \ ] &\\
    invariant &\Coloneqq formula &\\
    qualifier &\Coloneqq \mathbf{model} \mid \mathbf{ghost} &\\
    cardinality &\Coloneqq \mathbf{one} \mid \mathbf{lone} \mid \mathbf{some} \mid \mathbf{no} &\\
    \end{flalign*}
   
    \begin{flalign*}
    props &\Coloneqq acyclic \mid transitive \mid reflexive \mid irreflexive \mid symmetric &\\
          &\mid asymmetric \mid antisymmetric \mid total \mid functional \mid surjective &\\
          &\mid injective \mid bijective \mid complete \mid bijection  \mid preorder &\\
          &\mid equivalence \mid partialorder \mid totalorder &\\
    \\
    formula  &\Coloneqq \\
             &expr \subset expr               \tag{\text{subset}}            \label{syn:subset} &\\ 
    \mid \   &expr=expr                       \tag{\text{equality}}          \label{syn:equality} &\\
    \mid \   &\mathbf{some} \ expr            \tag{\text{non\_empty}}        \label{syn:nonempty} &\\
    \mid \   &\mathbf{one} \ expr             \tag{\text{exactly one}}       \label{syn:exactlyone} &\\
    \mid \   &\mathbf{lone} \ expr            \tag{\text{empty\ or\ one}}    \label{syn:emptyorone} &\\
    \mid \   &\mathbf{no} \ expr              \tag{\text{empty}}             \label{syn:empty} &\\
    \mid \   &\neg formula                    \tag{\text{negation}}          \label{syn:negation} &\\
    \mid \   &formula \wedge formula          \tag{\text{conjuction}}        \label{syn:conjunction} &\\
    \mid \   &formula \vee formula            \tag{\text{disjunction}}       \label{syn:disjunction} &\\
    \mid \   &formula \Rightarrow formula     \tag{\text{implication}}       \label{syn:implication} &\\
    \mid \   &\forall \ varDecls \mid formula \tag{\text{universal}}         \label{syn:universal} &\\
    \mid \   &\exists \ varDecls \mid formula \tag{\text{existential}}       \label{syn:existential} &\\
    \mid \   &\mathbf{intexpr} \; \{ {} = {} \mid {} < {} \mid {} > {} \} \; \mathbf{intexpr}
                                              \tag{\text{int \ comparison}}  \label{syn:intcomparison} &\\
    \\
    expr     &\Coloneqq \\
             &var                             \tag{\text{variable}}          \label{syn:variable} &\\ 
    \mid \   &expr=expr                       \tag{\text{equality}}          \label{syn:exprequality} &\\
    \mid \   &\sim expr                       \tag{\text{transpose}}         \label{syn:transpose} &\\
    \mid \   &\hat{\thickspace } expr         \tag{\text{clousure}}          \label{syn:clousure} &\\
    \mid \   &expr \cup expr                  \tag{\text{union}}             \label{syn:union} &\\
    \mid \   &expr \cap expr                  \tag{\text{intersection}}      \label{syn:intersection} &\\
    \mid \   &expr \setminus expr             \tag{\text{difference}}        \label{syn:difference} &\\
    \mid \   &expr \cdot expr                 \tag{\text{join}}              \label{syn:join} &\\
    \mid \   &expr \rightarrow expr           \tag{\text{product}}           \label{syn:product} &\\
    \mid \   &formula \, ? \; expr \; : \; expr  
                                              \tag{\text{conditional}}       \label{syn:conditional} &\\
    \mid \   &\{varDcls \mid formula \}       \tag{\text{comprehension}}     \label{syn:comprehension} &\\
    \mid \   &\pi (expr,intexpr^{\ast})       \tag{\text{projection}}        \label{syn:projection} &\\
    \mid \   &\mathbf{int2expr} \ (intexpr)   \tag{\text{int \ cast}}        \label{syn:intcast} \\
    \mid \   &\mathbf{univ}                   \tag{\text{universe}}          \label{syn:universe}
    \\
    intexpr  &\Coloneqq \\
             &integer                        \tag{\text{literal}}            \label{syn:literal} &\\ 
    \mid \   & expr                           \tag{\text{cardinality}}        \label{syn:cardinality} &\\
    \mid \   &\mathbf{sum} \; (e)            \tag{\text{sum}}                \label{syn:sum} &\\
    \mid \   &intexpr \; \{ + \mid \_ \mid \times \mid \div \} \; intexpr 
                                             \tag{\text{arithmetic}}         \label{syn:arithmetic} &\\
    \\
    varDecls &\Coloneqq (variable:expr)^{\ast} &\\
    variable &\Coloneqq identifier &\\
    \end{flalign*}

\section{Translation of The Theory of List}
\label{apx:translation}

In \figurename~\ref{fig:spec_tol}, \textit{theoryoflist.ecore} file and the partial instance given in \figurename~\ref{fig:inconsistency_detected} are being translated into KodKod API calls and then the result is interpreted to generated the \figurename~\ref{fig:reasoning_result}. The following listings are taken from \AIE's log file running this scenario and showing how \AIE constructs a KodKod problem from given metamodel and partial model to detect inconsistency (UNSAT) and to complete partial model (SAT).

\begin{enumerate}
    \item The universe of Theory of List instance (cf. Listing~\ref{lst:universe})
    \item Bounds for Unary Relations (cf. Listing~\ref{lst:unarybounds})
    \item Bounds for \texttt{Internal} Binary Relations (cf. Listing~\ref{lst:internalbinarybounds})
    \item Bounds for \texttt{User}'s Binary Relations (cf. Listing~\ref{lst:userbinarybounds})
    \item Generated Formulas (cf. Listing~\ref{lst:formulas})
    \item The Outcome and the statistics (cf. Listing~\ref{lst:stats})
    \item The Generated Model (cf. Listings~\ref{lst:modelunary} and~ \ref{lst:modelbinary})
\end{enumerate}

\lstset{basicstyle=\small, breakatwhitespace=true, breaklines=true}
\lstset{numbers=left,xleftmargin=2em,framexleftmargin=1.5em}

\begin{lstlisting}[float=*t, firstnumber=auto, firstline=1, lastline=1, label={lst:universe}, caption=Universe]
[EnginedVehicle84, EnginedVehicle184, TruckList729, Vehicle_3, TruckList848, TruckList400, Nil, EString_1, TruckList610, EString_3, TruckList, 3, 5, Vehicle, 1, EnginedVehicle, List_6, CarList, List, Object, NonEnginedVehicle, BicycleList, Memory, EnginedVehicle284, FORD F-150 XLT, String_0, CarList493, EInt, 4, 2, EString, List_5, Memory46]
\end{lstlisting}

\begin{lstlisting}[float=*t, firstnumber=last, firstline=3, lastline=34, label={lst:unarybounds}, caption=Bounds for Unary Relations]
relation bounds:
 BicycleList: [[], [[List_6], [List_5]]]
 Class: [[[Nil], [TruckList], [Vehicle], [EnginedVehicle], [CarList], [List], [Object], [NonEnginedVehicle], [BicycleList], [Memory], [EInt], [EString]]]
 Nil<EnginedVehicle>: [[], [[List_6], [List_5]]]
 Class<EnginedVehicle>: [[[EnginedVehicle]]]
 EnginedVehicle: [[[EnginedVehicle84], [EnginedVehicle184], [EnginedVehicle284]], [[EnginedVehicle84], [EnginedVehicle184], [Vehicle_3], [EnginedVehicle284]]]
 Class<Vehicle>: [[[Vehicle]]]
 Nil<NonEnginedVehicle>: [[], [[List_6], [List_5]]]
 List<? extends Vehicle>: [[[TruckList729], [TruckList848], [TruckList400], [TruckList610], [CarList493]], [[TruckList729], [TruckList848], [TruckList400], [TruckList610], [List_6], [CarList493], [List_5]]]
 List: [[[TruckList729], [TruckList848], [TruckList400], [TruckList610], [CarList493], [List_5]], [[TruckList729], [TruckList848], [TruckList400], [TruckList610], [List_6], [CarList493], [List_5]]]
 Class<CarList>: [[[CarList]]]
 Class<Nil>: [[[Nil]]]
 "FORD F-150 XLT": [[[FORD F-150 XLT]]]
 Memory: [[[Memory46]]]
 List<EnginedVehicle>: [[[TruckList729], [TruckList848], [TruckList400], [TruckList610], [CarList493]], [[TruckList729], [TruckList848], [TruckList400], [TruckList610], [List_6], [CarList493], [List_5]]]
 Class<List>: [[[List]]]
 Vehicle: [[[EnginedVehicle84], [EnginedVehicle184], [EnginedVehicle284]], [[EnginedVehicle84], [EnginedVehicle184], [Vehicle_3], [EnginedVehicle284]]]
 Class<BicycleList>: [[[BicycleList]]]
 EString: [[[EString_1], [EString_3], [FORD F-150 XLT]], [[EString_1], [EString_3], [FORD F-150 XLT], [String_0]]]
 Class<EInt>: [[[EInt]]]
 List<NonEnginedVehicle>: [[], [[List_6], [List_5]]]
 Class<EString>: [[[EString]]]
 Object: [[[EnginedVehicle84], [EnginedVehicle184], [TruckList729], [TruckList848], [TruckList400], [TruckList610], [EnginedVehicle284], [CarList493], [List_5]], [[EnginedVehicle84], [EnginedVehicle184], [TruckList729], [Vehicle_3], [TruckList848], [TruckList400], [TruckList610], [List_6], [EnginedVehicle284], [CarList493], [List_5]]]
 Class<TruckList>: [[[TruckList]]]
 NonEnginedVehicle: [[], [[Vehicle_3]]]
 CarList: [[[CarList493]], [[List_6], [CarList493], [List_5]]]
 EInt: [[[3], [5], [1], [4], [2]]]
 TruckList: [[[TruckList729], [TruckList848], [TruckList400], [TruckList610]], [[TruckList729], [TruckList848], [TruckList400], [TruckList610], [List_6], [List_5]]]
 Class<NonEnginedVehicle>: [[[NonEnginedVehicle]]]
 Nil: [[], [[List_6], [List_5]]]
 Class<Memory>: [[[Memory]]]
 Class<Object>: [[[Object]]]
\end{lstlisting}

\begin{lstlisting}[float=*t, firstnumber=last, firstline=35, lastline=38, label={lst:internalbinarybounds}, caption=Lower and Upper Bounds for \texttt{Internal} Binary Relations]
 extends: [[[Vehicle, EnginedVehicle], [Vehicle, NonEnginedVehicle], [List, Nil], [List, TruckList], [List, CarList], [List, BicycleList], [Object, Nil], [Object, TruckList], [Object, Vehicle], [Object, EnginedVehicle], [Object, CarList], [Object, List], [Object, NonEnginedVehicle], [Object, BicycleList]]]
 super: [[[Nil, List], [Nil, Object], [TruckList, List], [TruckList, Object], [Vehicle, Object], [EnginedVehicle, Vehicle], [EnginedVehicle, Object], [CarList, List], [CarList, Object], [List, Object], [NonEnginedVehicle, Vehicle], [NonEnginedVehicle, Object], [BicycleList, List], [BicycleList, Object]]]
 lists: [[[Memory46, TruckList729], [Memory46, TruckList848], [Memory46, TruckList400], [Memory46, TruckList610], [Memory46, CarList493]], [[Memory46, TruckList729], [Memory46, TruckList848], [Memory46, TruckList400], [Memory46, TruckList610], [Memory46, List_6], [Memory46, CarList493], [Memory46, List_5]]]
 class: [[[EnginedVehicle84, Vehicle], [EnginedVehicle84, EnginedVehicle], [EnginedVehicle84, Object], [EnginedVehicle184, Vehicle], [EnginedVehicle184, EnginedVehicle], [EnginedVehicle184, Object], [TruckList729, TruckList], [TruckList729, List], [TruckList729, Object], [TruckList848, TruckList], [TruckList848, List], [TruckList848, Object], [TruckList400, TruckList], [TruckList400, List], [TruckList400, Object], [EString_1, EString], [TruckList610, TruckList], [TruckList610, List], [TruckList610, Object], [EString_3, EString], [3, EInt], [5, EInt], [1, EInt], [EnginedVehicle284, Vehicle], [EnginedVehicle284, EnginedVehicle], [EnginedVehicle284, Object], [FORD F-150 XLT, EString], [CarList493, CarList], [CarList493, List], [CarList493, Object], [4, EInt], [2, EInt], [List_5, List], [List_5, Object], [Memory46, Memory]], [[EnginedVehicle84, Vehicle], [EnginedVehicle84, EnginedVehicle], [EnginedVehicle84, Object], [EnginedVehicle184, Vehicle], [EnginedVehicle184, EnginedVehicle], [EnginedVehicle184, Object], [TruckList729, TruckList], [TruckList729, List], [TruckList729, Object], [Vehicle_3, Vehicle], [Vehicle_3, EnginedVehicle], [Vehicle_3, Object], [Vehicle_3, NonEnginedVehicle], [TruckList848, TruckList], [TruckList848, List], [TruckList848, Object], [TruckList400, TruckList], [TruckList400, List], [TruckList400, Object], [EString_1, EString], [TruckList610, TruckList], [TruckList610, List], [TruckList610, Object], [EString_3, EString], [3, EInt], [5, EInt], [1, EInt], [List_6, Nil], [List_6, TruckList], [List_6, CarList], [List_6, List], [List_6, Object], [List_6, BicycleList], [EnginedVehicle284, Vehicle], [EnginedVehicle284, EnginedVehicle], [EnginedVehicle284, Object], [FORD F-150 XLT, EString], [String_0, EString], [CarList493, CarList], [CarList493, List], [CarList493, Object], [4, EInt], [2, EInt], [List_5, Nil], [List_5, TruckList], [List_5, CarList], [List_5, List], [List_5, Object], [List_5, BicycleList], [Memory46, Memory]]]
\end{lstlisting}

\begin{lstlisting}[float=*t, firstnumber=last, firstline=39, lastline=43, label={lst:userbinarybounds}, caption=Lower and Upper Bounds for \texttt{User}'s Binary Relations]
 cdr: [[[TruckList729, TruckList400], [TruckList848, TruckList400], [TruckList610, TruckList729], [CarList493, TruckList848]], [[TruckList729, TruckList729], [TruckList729, TruckList848], [TruckList729, TruckList400], [TruckList729, TruckList610], [TruckList729, List_6], [TruckList729, CarList493], [TruckList729, List_5], [TruckList848, TruckList729], [TruckList848, TruckList848], [TruckList848, TruckList400], [TruckList848, TruckList610], [TruckList848, List_6], [TruckList848, CarList493], [TruckList848, List_5], [TruckList400, TruckList729], [TruckList400, TruckList848], [TruckList400, TruckList400], [TruckList400, TruckList610], [TruckList400, List_6], [TruckList400, CarList493], [TruckList400, List_5], [TruckList610, TruckList729], [TruckList610, TruckList848], [TruckList610, TruckList400], [TruckList610, TruckList610], [TruckList610, List_6], [TruckList610, CarList493], [TruckList610, List_5], [List_6, TruckList729], [List_6, TruckList848], [List_6, TruckList400], [List_6, TruckList610], [List_6, List_6], [List_6, CarList493], [List_6, List_5], [CarList493, TruckList729], [CarList493, TruckList848], [CarList493, TruckList400], [CarList493, TruckList610], [CarList493, List_6], [CarList493, CarList493], [CarList493, List_5], [List_5, TruckList729], [List_5, TruckList848], [List_5, TruckList400], [List_5, TruckList610], [List_5, List_6], [List_5, CarList493], [List_5, List_5]]]
 car: [[[TruckList729, EnginedVehicle84], [TruckList848, EnginedVehicle84], [TruckList400, EnginedVehicle184], [TruckList610, EnginedVehicle184], [CarList493, EnginedVehicle184]], [[TruckList729, EnginedVehicle84], [TruckList729, EnginedVehicle184], [TruckList729, Vehicle_3], [TruckList729, EnginedVehicle284], [TruckList848, EnginedVehicle84], [TruckList848, EnginedVehicle184], [TruckList848, Vehicle_3], [TruckList848, EnginedVehicle284], [TruckList400, EnginedVehicle84], [TruckList400, EnginedVehicle184], [TruckList400, Vehicle_3], [TruckList400, EnginedVehicle284], [TruckList610, EnginedVehicle84], [TruckList610, EnginedVehicle184], [TruckList610, Vehicle_3], [TruckList610, EnginedVehicle284], [List_6, EnginedVehicle84], [List_6, EnginedVehicle184], [List_6, Vehicle_3], [List_6, EnginedVehicle284], [CarList493, EnginedVehicle84], [CarList493, EnginedVehicle184], [CarList493, Vehicle_3], [CarList493, EnginedVehicle284], [List_5, EnginedVehicle84], [List_5, EnginedVehicle184], [List_5, Vehicle_3], [List_5, EnginedVehicle284]]]
 eq: [[], [[TruckList729, TruckList729], [TruckList729, TruckList848], [TruckList729, TruckList400], [TruckList729, TruckList610], [TruckList729, List_6], [TruckList729, CarList493], [TruckList729, List_5], [TruckList848, TruckList729], [TruckList848, TruckList848], [TruckList848, TruckList400], [TruckList848, TruckList610], [TruckList848, List_6], [TruckList848, CarList493], [TruckList848, List_5], [TruckList400, TruckList729], [TruckList400, TruckList848], [TruckList400, TruckList400], [TruckList400, TruckList610], [TruckList400, List_6], [TruckList400, CarList493], [TruckList400, List_5], [TruckList610, TruckList729], [TruckList610, TruckList848], [TruckList610, TruckList400], [TruckList610, TruckList610], [TruckList610, List_6], [TruckList610, CarList493], [TruckList610, List_5], [List_6, TruckList729], [List_6, TruckList848], [List_6, TruckList400], [List_6, TruckList610], [List_6, List_6], [List_6, CarList493], [List_6, List_5], [CarList493, TruckList729], [CarList493, TruckList848], [CarList493, TruckList400], [CarList493, TruckList610], [CarList493, List_6], [CarList493, CarList493], [CarList493, List_5], [List_5, TruckList729], [List_5, TruckList848], [List_5, TruckList400], [List_5, TruckList610], [List_5, List_6], [List_5, CarList493], [List_5, List_5]]]
 name: [[[EnginedVehicle84, EString_3], [EnginedVehicle184, EString_1]], [[EnginedVehicle84, EString_1], [EnginedVehicle84, EString_3], [EnginedVehicle84, FORD F-150 XLT], [EnginedVehicle84, String_0], [EnginedVehicle184, EString_1], [EnginedVehicle184, EString_3], [EnginedVehicle184, FORD F-150 XLT], [EnginedVehicle184, String_0], [Vehicle_3, EString_1], [Vehicle_3, EString_3], [Vehicle_3, FORD F-150 XLT], [Vehicle_3, String_0], [EnginedVehicle284, EString_1], [EnginedVehicle284, EString_3], [EnginedVehicle284, FORD F-150 XLT], [EnginedVehicle284, String_0]]]
 vehicles: [[[Memory46, EnginedVehicle84], [Memory46, EnginedVehicle184], [Memory46, EnginedVehicle284]], [[Memory46, EnginedVehicle84], [Memory46, EnginedVehicle184], [Memory46, Vehicle_3], [Memory46, EnginedVehicle284]]]
\end{lstlisting}

\begin{lstlisting}[float=*t, firstnumber=last, firstline=47, lastline=100, label={lst:formulas}, caption=Generated Formulas]
List = (List<? extends Vehicle> + Nil) 
(all n: univ | Class<Nil> in (n.class) <=> n in Nil) 
no (Memory & Object) 
Vehicle = (EnginedVehicle + NonEnginedVehicle) 
(all e: univ | Class<EnginedVehicle> in (e.class) <=> e in EnginedVehicle) 
(Memory.*(lists + vehicles)) = (Memory + Object) 
(all m: Memory.lists | lone (m.~(lists + vehicles))) 
no (NonEnginedVehicle & EnginedVehicle) 
(all l: List | lone (l.car)) 
Object = (Vehicle + List) 
cdr in ((List<EnginedVehicle> -> List<EnginedVehicle>) + (List<NonEnginedVehicle> -> List<NonEngin...
(all l: List | lone (l.cdr)) 
(all a: Vehicle, b: Vehicle | !(a = b) => !((a.name) = (b.name))) 
no (Nil.cdr) 
(all o: univ | Class<Object> in (o.class) <=> o in Object) 
no (CarList & Nil<EnginedVehicle>) 
List<? extends Vehicle> = (List<EnginedVehicle> + List<NonEnginedVehicle>) 
(all b: univ | Class<BicycleList> in (b.class) <=> b in BicycleList) 
(all m: Memory.vehicles | lone (m.~(lists + vehicles))) 
no (TruckList & Nil<EnginedVehicle>) 
List<EnginedVehicle> = CarList + TruckList + Nil<EnginedVehicle> 
one Nil 
no (BicycleList & Nil<NonEnginedVehicle>) 
no (List<EnginedVehicle> & List<NonEnginedVehicle>) 
no (Nil.car) 
all a: List, b: List | a in b.eq <=> (a.car = b.car and a.cdr) in ((b.cdr).eq and a.class = b.class)
(all l: univ | Class<List> in (l.class) <=> l in List) 
(all l: List | l in (l.eq)) 
some vehicles 
(all l: List - Nil | some (l.cdr) and some (l.car)) 
Nil = (Nil<NonEnginedVehicle> + Nil<EnginedVehicle>) 
(all e: univ | Class<EInt> in (e.class) <=> e in EInt) 
eq in ((List<NonEnginedVehicle> -> List<NonEnginedVehicle>) + (List<EnginedVehicle> -> List<Engine...
(all t: univ | Class<TruckList> in (t.class) <=> t in TruckList) 
(all l: List | !(l in (l.^cdr))) 
lists in (Memory -> List<? extends Vehicle>) 
no (Vehicle & List) 
one Memory 
vehicles in (Memory -> Vehicle) 
(all v: univ | Class<Vehicle> in (v.class) <=> v in Vehicle) 
(all c: univ | Class<CarList> in (c.class) <=> c in CarList) 
List<NonEnginedVehicle> = (BicycleList + Nil<NonEnginedVehicle>) 
(all l: List | Nil in (l.*cdr)) 
(all v: Object - List | some (v.~car)) 
no (Nil<EnginedVehicle> & Nil<NonEnginedVehicle>) 
(all n: univ | Class<NonEnginedVehicle> in (n.class) <=> n in NonEnginedVehicle) 
no (TruckList & CarList) 
some lists 
(all e: univ | Class<EString> in (e.class) <=> e in EString) 
car in ((List<NonEnginedVehicle> -> NonEnginedVehicle) + (List<EnginedVehicle> -> EnginedVehicle)) 
name in (Vehicle -> EString) 
(some v: Vehicle | all v': Vehicle |  (v'.name) = "FORD F-150 XLT" <=>  v' = v) 
(all m: univ | Class<Memory> in (m.class) <=> m in Memory) 
(all v: Vehicle | one (v.name))
\end{lstlisting}

\begin{lstlisting}[float=*t, firstnumber=last, firstline=102, lastline=109, label={lst:stats}, caption=The Outcome and the statistics of the reasoning process]
---OUTCOME---
SATISFIABLE

---STATS---
p cnf 2646 6717
primary variables: 178
translation time: 149 ms
solving time: 10 ms
\end{lstlisting}

\begin{lstlisting}[float=*t, firstnumber=last, firstline=111, lastline=142, label={lst:modelunary}, caption=The Generated Model (Unary Relations)]
---MODEL---
Nil<V>:         List_5
List<E>:        TruckList610, TruckList400, List_5, CarList493, List_6, TruckList848, TruckList729
Class<Nil>:     Nil
CarList:        CarList493
Class<EInt>:    EInt
Vehicle:        EnginedVehicle184, EnginedVehicle284, EnginedVehicle84
List<EnginedVehicle>: TruckList610, TruckList400, List_5, CarList493, List_6, TruckList848, TruckList729
Class<Vehicle>: Vehicle
NonEnginedVehicle: 
Class<Memory>:  Memory
Object:         TruckList610, TruckList400, List_5, EnginedVehicle184, EnginedVehicle284, CarList493, List_6, EnginedVehicle84, TruckList848, TruckList729
TruckList:      TruckList610, TruckList400, List_6, TruckList848, TruckList729
Nil<NonEnginedVehicle>: 
EString:        EString_1, EString_3, FORD F-150 XLT
EInt:           2, 1, 4, 3, 5
Class<List>:    List
Class<Object>:  Object
Class:          Nil, EString, NonEnginedVehicle, CarList, Vehicle, BicycleList, EInt, List, Memory, EnginedVehicle, Object, TruckList
EnginedVehicle: EnginedVehicle184, EnginedVehicle284, EnginedVehicle84
Class<CarList>: CarList
Class<TruckList>: TruckList
List<? extends Vehicle>: TruckList610, TruckList400, List_5, CarList493, List_6, TruckList848, TruckList729
Nil<EnginedVehicle>: List_5
List<NonEnginedVehicle>: 
BicycleList:    
"FORD F-150 XLT": FORD F-150 XLT
Class<EString>: EString
Class<NonEnginedVehicle>: NonEnginedVehicle
Class<EnginedVehicle>: EnginedVehicle
Class<BicycleList>: BicycleList
Memory:         Memory46
\end{lstlisting}

\begin{lstlisting}[float=*t, firstnumber=last, firstline=144, lastline=177, label={lst:modelbinary}, caption=The Generated Model (Binary Relations)]
eq (TruckList610, TruckList610)
eq (TruckList848, TruckList729)
eq (TruckList729, TruckList729)
eq (TruckList400, TruckList400)
eq (TruckList729, TruckList848)
eq (List_5, List_5)
eq (CarList493, CarList493)
eq (List_6, List_6)
eq (TruckList848, TruckList848)
car (TruckList400, EnginedVehicle184)
car (List_6, EnginedVehicle284)
car (TruckList610, EnginedVehicle184)
car (CarList493, EnginedVehicle184)
car (TruckList848, EnginedVehicle84)
car (TruckList729, EnginedVehicle84)
vehicles (Memory46, EnginedVehicle184)
vehicles (Memory46, EnginedVehicle84)
vehicles (Memory46, EnginedVehicle284)
lists (Memory46, CarList493)
lists (Memory46, List_5)
lists (Memory46, List_6)
lists (Memory46, TruckList610)
lists (Memory46, TruckList400)
lists (Memory46, TruckList729)
lists (Memory46, TruckList848)
name (EnginedVehicle284, FORD F-150 XLT)
name (EnginedVehicle184, EString_1)
name (EnginedVehicle84, EString_3)
cdr (List_6, List_5)
cdr (TruckList848, TruckList400)
cdr (TruckList729, TruckList400)
cdr (TruckList400, List_6)
cdr (TruckList610, TruckList729)
cdr (CarList493, TruckList848)
\end{lstlisting}

\end{document}